\documentclass{PoS}
\title{Towards global fits in EFT's and New Physics implications}

\ShortTitle{Towards global fits in EFT's and New Physics implications}

\author{\speaker{Emma Slade}\\
Rudolf Peierls Centre for Theoretical Physics, University of Oxford, \\
 Clarendon Laboratory, Parks Road, Oxford OX1 3PU, United Kingdom \\
        E-mail: \email{emma.slade@physics.ox.ac.uk}}

\abstract{I discuss recent progress on fits to dimension-six operators in the Standard Model Effective Theory (SMEFT).
I focus on the top quark sector of the SMEFT, as well as the theoretical advances made in computing SMEFT effects through to next-to-leading order in QCD and the use of these calculations in global fits.
I also discuss fits performed to the Higgs and electroweak sectors of the SMEFT and the possibility for performing global fits to multiple sectors simultaneously.}

\FullConference{7th Annual Conference on Large Hadron Collider Physics - LHCP2019\\
		20-25 May, 2019\\
		Puebla, Mexico}

\begin{document}

\paragraph{Introduction}
A powerful framework to identify and parametrise 
deviations with respect to the Standard Model (SM) predictions in a model-independent way is
the Standard Model Effective Field Theory
(SMEFT)~\cite{Weinberg:1979sa,Buchmuller:1985jz,Grzadkowski:2010es} - see~\cite{Brivio:2017vri} for a recent review.
In the SMEFT, physics beyond the SM which manifests at high scales $E\simeq \Lambda$ is parameterised, at the accessible scale $E\ll \Lambda$, in terms of higher-dimensional operators built up from the SM fields and symmetries.
This technique allows one to construct complete bases of 
independent operators at any mass dimension, which can then 
 be matched to ultraviolet-complete theories.
The resulting Lagrangian then admits a power expansion 

\begin{equation}
\label{eq:smeftlagrangian}
\mathcal{L}_{\rm SMEFT}=\mathcal{L}_{\rm SM} + \sum_i^{N_{d6}} \frac{c_i}{\Lambda^2}\mathcal{O}_i^{(6)} +
\sum_j^{N_{d8}} \frac{b_j}{\Lambda^4}\mathcal{O}_j^{(8)} + \ldots \, ,
\end{equation}
where $\mathcal{L}_{\rm SM}$ is the SM Lagrangian,  $c_i$ are the (unknown) Wilson coefficients, and
$\{\mathcal{O}_i^{(6)}\}$ and $\{\mathcal{O}_j^{(8)}\}$ stand for
the elements of the
operator basis of mass-dimension $d=6$ and $d=8$,
respectively.

The number of operators that enter at each mass dimension is known~\cite{Henning:2015alf,Lehman:2015coa}, as are the complete bases of operators at dimensions 5-7~\cite{PhysRevLett.43.1566, Leung1986,Buchmuller:1985jz,Grzadkowski:2010es,Lehman:2014jma,Henning:2015alf}. 
Operators with $d=5$ and $d=7$, which violate lepton and/or baryon number
conservation~\cite{Degrande:2012wf,Kobach:2016ami}, and are usually not considered in fits to the SMEFT using LHC data. The dimension-6 terms are therefore the leading deviation from the SM.

In general, the effects of the dimension-6 operators can be written as follows:
\begin{equation}
\label{eq:smeftXsecInt}
\sigma=\sigma_{\rm SM} + \sum_i^{N_{d6}}
\kappa_i \frac{c_i}{\Lambda^2} +
\sum_{i,j}^{N_{d6}}  \widetilde{\kappa}_{ij} \frac{c_ic_j}{\Lambda^4}  \, ,
\end{equation}
where $\sigma_{\rm SM}$ indicates
the SM prediction and $c_i$ are the Wilson coefficients we wish to constrain.
The $\mathcal{O}(\Lambda^{-2})$ corrections to the SM 
cross-sections represent formally the dominant correction.
The third term representing $\mathcal{O}(\Lambda^{-4})$ effects, and are from the
squared amplitudes of the SMEFT operators.
In principle, this term may not need to be included, depending on whether the truncation at
$\mathcal{O}(\Lambda^{-2})$ order is done at the Lagrangian or the cross-section level, but in practice there are often
valid reasons to include them in the calculation.

In general, the operators run with the scale, meaning the coefficients, $c_i$, will depend on the typical momentum transfer of the process. 
This dependence can be computed using the renormalisation group equations (RGE), which at dimension-6 are fully known~\cite{Jenkins:2013zja,Jenkins:2013wua,Alonso:2013hga,Grojean:2013kd,Alonso:2014zka,Ghezzi:2015vva}. 
One can typically ignore operator-mixing effects when focusing on processes with a similar energy scale, e.g. $E \simeq m_t$. Additionally, the inclusion of NLO QCD corrections will reduce this scale dependence, making the RG effects less significant~\cite{Maltoni:2016yxb,Deutschmann:2017qum}.

\paragraph{The top quark sector of the SMEFT}
An important aspect of any SMEFT analysis is the need to include
all relevant operators that contribute
to the processes whose data is used as input
to the fit.
Only in this way can the SMEFT retain its
model and basis independence.
However, unless specific scenarios are adopted, the number of
non-redundant operators  becomes unfeasibly large:
59 for one generation of fermions \cite{Grzadkowski:2010es} and 2499 for
three~\cite{Alonso:2013hga} at dimension-6.
This implies that a global SMEFT fit
will have to explore a huge
parameter space with potentially a large number of flat (degenerate) directions.

\begin{figure}[t]
  \begin{center}
\includegraphics[scale=0.45]{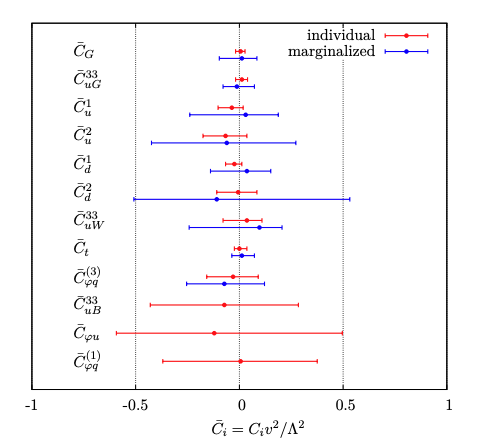}
    \caption{\small The 95\% CL bounds on the degrees of freedom
      included the TopFitter analysis, both in the marginalised  
      and in the individual fit cases. 
      The definitions of the operators is given in~\cite{Buckley:2015lku}. 
      Figure from~\cite{Buckley:2015lku}.
      \label{fig:TopFitter-bounds}
  }
  \end{center}
\end{figure}

Looking initially at just the top quark sector of the SMEFT, there are a couple of groups who have performed fits to the subset of operators relevant to top physics.
TopFitter~\cite{Buckley:2015nca,Buckley:2015lku} use parton-level measurements of $t\bar{t}$-pair production, single-top production and $t\bar{t} \gamma/Z$ production from the LHC run I and II and Tevatron, with a total of 227 measurements.
Fitting is performed using the \texttt{PROFESSOR}~\cite{Buckley:2009bj} framework, with the SMEFT corrections computed at tree-level and neglecting $\mathcal{O}(\Lambda^{-4})$ effects.
The result of the TopFitter analysis is shown in Fig.~\ref{fig:TopFitter-bounds}, both at the marginalised and individual fit levels. 
TopFitter have recently~\cite{Brown:2019pzx} extended their analysis to include particle-level measurements using run II data from the LHC.

The SMEFiT collaboration~\cite{Hartland:2019bjb} follows the strategy
of the LHC Top Quark Working Group 
note~\cite{AguilarSaavedra:2018nen}.
They adopt the Minimal
Flavour Violation (MFV) hypothesis~\cite{DAmbrosio:2002vsn} in the quark
sector as the baseline scenario.
They additionally impose a $U(2)_q\times U(2)_u \times U(2)_d$ flavour symmetry in the first two generations.
They therefore consider 34 degrees of freedom, which are constrained by $t\bar{t}$-pair production, single-top production, as well as $t\bar{t}$ and single-top associated production, with a total of 103 measurements from the LHC run II.
SMEFiT compute all the SMEFT contributions at $\mathcal{O}(\Lambda^{-4})$  with NLO QCD corrections where available~\cite{Degrande:2018fog,Maltoni:2016yxb,Zhang:2016omx,Degrande:2014tta,Franzosi:2015osa,Bylund:2016phk,Durieux:2018tev}, and the SM calculations at NNLO for available processes. 

\begin{figure}[t]
  \begin{center}
\includegraphics[scale=0.4]{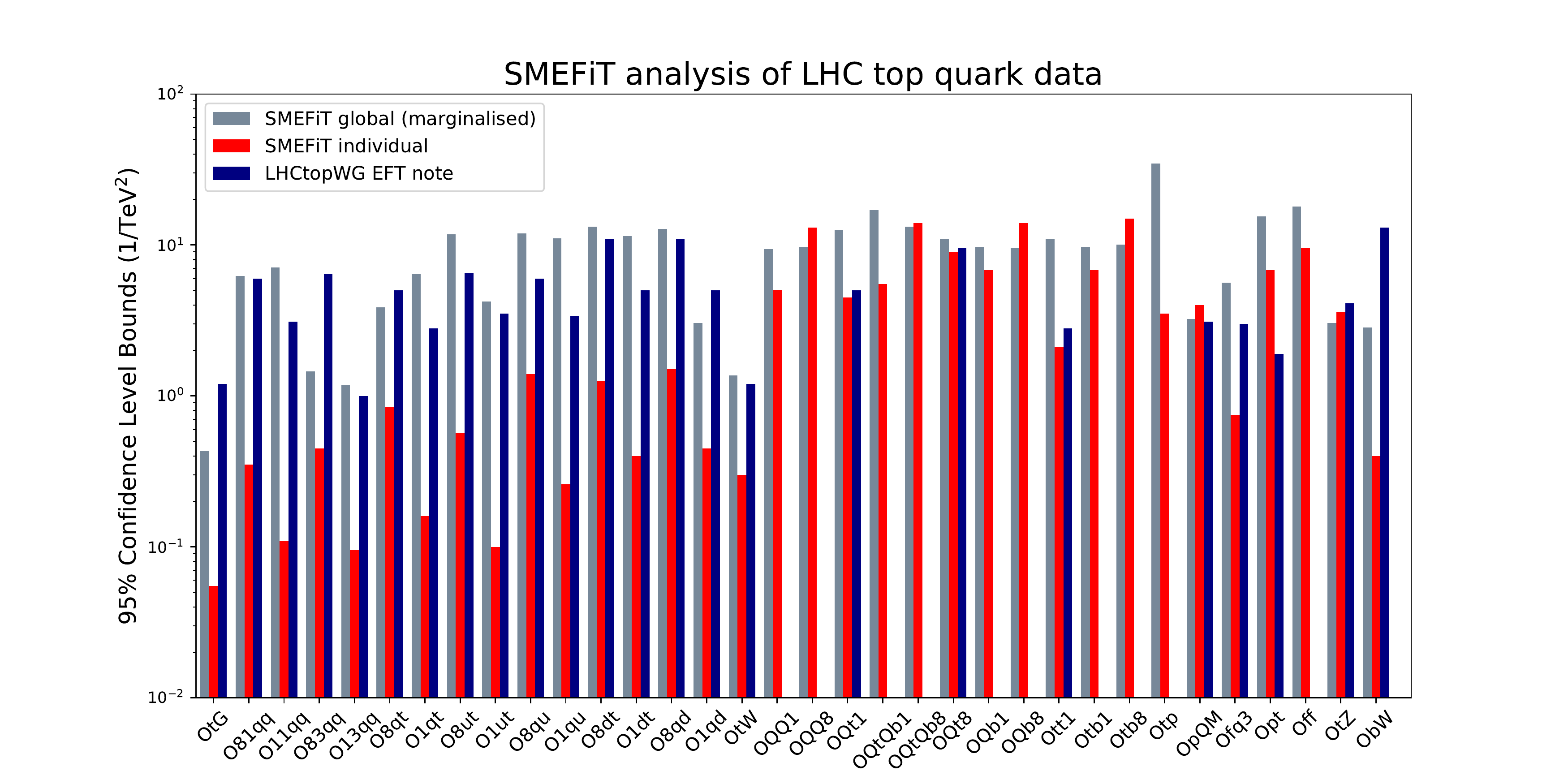}
    \caption{\small The 95\% CL bounds on the 34 degrees of freedom
      included in SMEFiT, both in the marginalised
      and in the individual fit cases, with the
      bounds reported in the LHC Top WG EFT note~\cite{AguilarSaavedra:2018nen}.
The definitions of the operators is given in Ref.~\cite{Hartland:2019bjb}. Figure from Ref.~\cite{Hartland:2019bjb}.
      \label{fig:SMEFiT-bounds}
  }
  \end{center}
\end{figure}

In Fig.~\ref{fig:SMEFiT-bounds} we show the bounds computed using the SMEFiT methodology -- as different flavour assumptions are used by TopFitter and SMEFiT, one cannot directly compare the bounds obtained by the two groups for all operators. 
In general, within finite-size uncertainties, the bounds from individual fits will be much tighter than the marginalised bounds because in the former, correlations between the degrees of freedom are neglected, which may be very large. 
Care must therefore be taken when using individual bounds, as they will in general be unrealistically tight.

\begin{figure}[h!]
\centering
\includegraphics[scale=0.4]{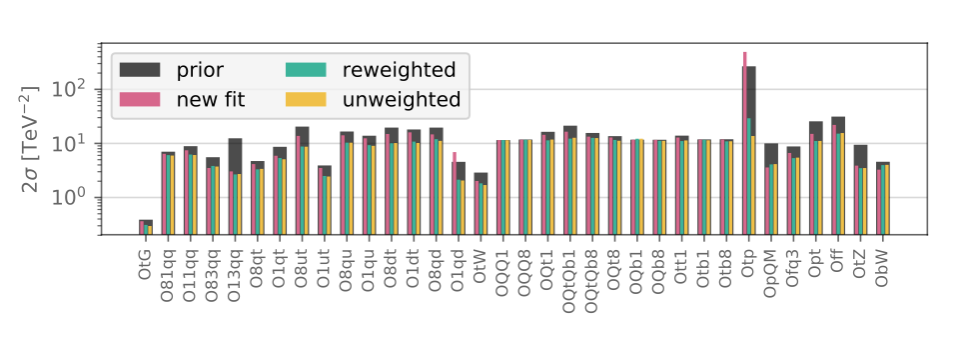}
\caption{\small The 95\% CL bounds for
    the $N_{\rm op}=34$ Wilson coefficients considered in the SMEFiT reweighting analysis 
    of the top quark sector. Figure from~\cite{vanBeek:2019evb}.
\label{fig:two_sigma_bounds_t_channel} }
\end{figure}

SMEFiT have recently reported~\cite{vanBeek:2019evb} on the applicability of the Bayesian reweighting technique developed for fitting Parton Distribution Functions~\cite{Ball:2010gb,Ball:2011gg}. 
This method has two advantages in comparison to a new fit to an extended set of data: first, it is essentially instantaneous, and second, it can be carried out without needing access to the original SMEFT fitting code.
They find that, under well-defined conditions, the results obtained with reweighting all the single-top $t$-channel data are equivalent to those obtained with a new fit to the extended set of data.
We show in Fig.~\ref{fig:two_sigma_bounds_t_channel}  the prior results without any $t$- or $s$-channel single-top 
    production data included with those after the $t$-channel measurements 
    have been added either by reweighting or by performing a new fit. 
    Within finite-size uncertainties, excellent agreement is found between the reweighted and new fits.
    The reweighting procedure is not restricted to the SMEFiT analysis; any Monte Carlo (MC) SMEFT fit, such as Sfitter~\cite{Biekotter:2018rhp,Moutafis:2019wbp}, should be able to use reweighting in order to gain an understanding of the impact of new data.

\begin{figure}[t]
  \begin{center}
\includegraphics[scale=0.35]{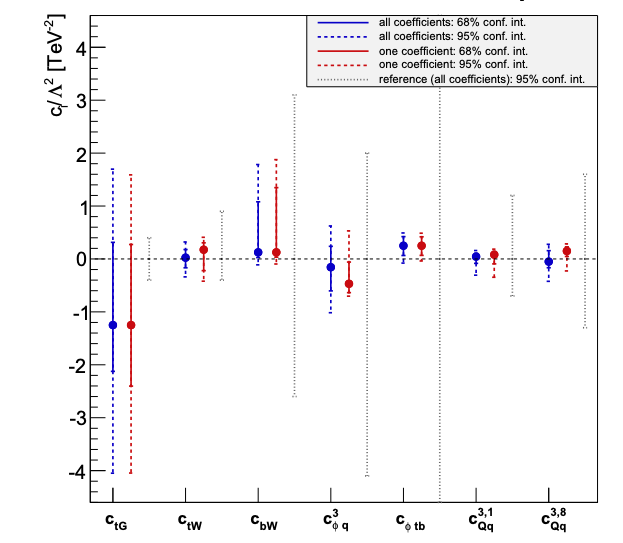}
    \caption{\small The 68\% and 95\% CL bounds on the degrees of freedom
      included in Sfitter top analysis, both in the marginalised
      and in the individual fit cases.
The definitions of the operators is given in Ref.~\cite{Moutafis:2019wbp}. Figure from Ref.~\cite{Moutafis:2019wbp}.
      \label{fig:sfitter-top}
  }
  \end{center}
\end{figure}

The Sfitter collaboration have also performed a fit to the top quark sector~\cite{Moutafis:2019wbp}. They look at the coefficients that are constrained
by the production or decay of a top quark, computing all available SMEFT measurements to NLO QCD where available.
As with the SMEFiT methodology, Sfitter represents the probability density in the space of Wilson coefficients as an ensemble of MC replicas.
By using a MC methodology, one avoids making any assumptions about the underlying probability distributions of the Wilson coefficients, and does not require Gaussian experimental uncertainties.
We show in Fig.~\ref{fig:sfitter-top} the bounds obtained by the Sfitter analysis at the 68\% and 95\% CL.

\begin{figure}[t!]
  \begin{center}
\includegraphics[scale=0.5]{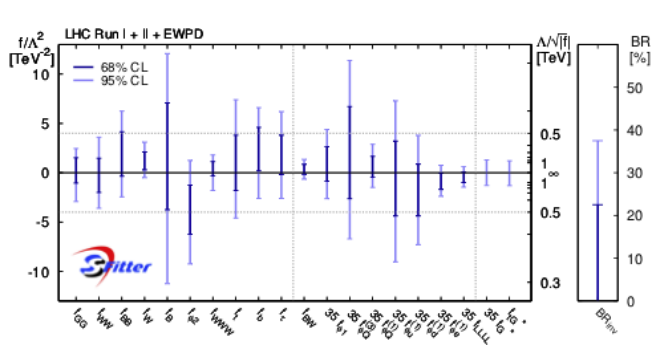}
    \caption{\small The 68\% and  95\% CL bounds for individual Wilson coefficients
      included in the Sfitter electroweak and Higgs analysis.
The definitions of the operators given in Ref.~\cite{Biekotter:2018rhp}. Figure from Ref.~\cite{Biekotter:2018rhp}.
      \label{fig:Sfitter-bounds}
  }
  \end{center}
\end{figure}

\begin{figure}[h!]
  \begin{center}
\includegraphics[scale=0.45]{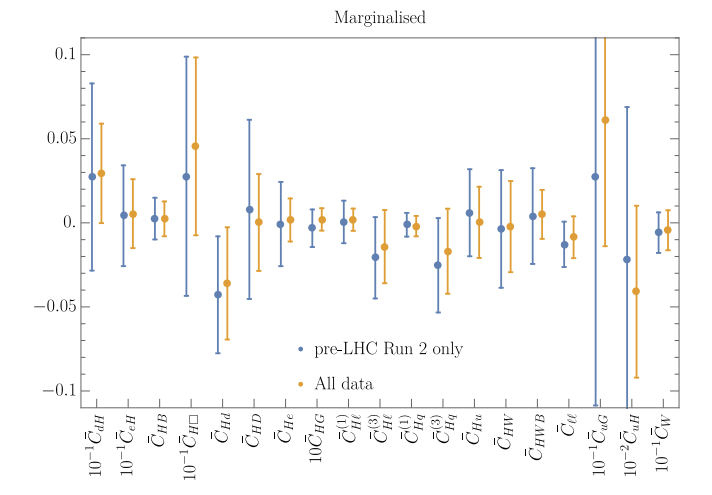}
    \caption{\small The 68\% and  95\% CL bounds for marginalised Wilson coefficients
      included in Ref.~\cite{Ellis:2018gqa}.
The definitions of the operators given in Ref.~\cite{Ellis:2018gqa}. Figure from Ref.~\cite{Ellis:2018gqa}.
      \label{fig:ellis-bounds}
  }
  \end{center}
\end{figure}

\begin{figure}[h!]
  \begin{center}
\includegraphics[scale=0.55]{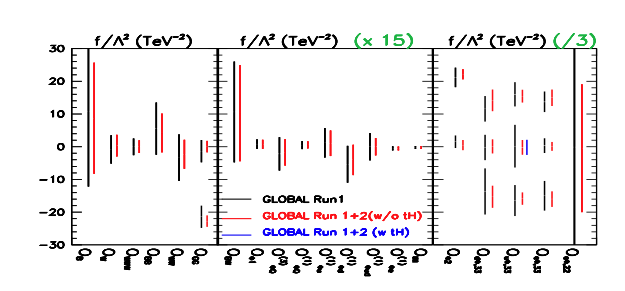}
    \caption{\small The 95\% CL bounds for marginalised Wilson coefficients
      included in Ref.~\cite{Almeida:2018cld}.
The definitions of the operators given in Ref.~\cite{Almeida:2018cld}. Figure from Ref.~\cite{Almeida:2018cld}.
      \label{fig:eboli-bounds}
  }
  \end{center}
\end{figure}

\paragraph{Towards global fits in the SMEFT}
In order to be able to perform global fits to the SMEFT, one needs to be able to combine data from different sectors of the SMEFT, such as the top and Higgs sectors.
Several groups have performed fits to the electroweak precision observables (EWPO) or Higgs data, such as Gfitter~\cite{Baak:2014ora}, Zfitter~\cite{Akhundov:2013ons} and HEPFit~\cite{deBlas:2017wmn}.
The Sfitter collaboration have recently~\cite{Biekotter:2018rhp} performed non-marginalised fits to the EWPO and Higgs data simultaneously, including $\mathcal{O}(\Lambda^{-4})$ effects in the SMEFT at LO in the SMEFT.
In Fig.~\ref{fig:Sfitter-bounds} we show the results of individual fits to the operators considered in the Sfitter analysis at the 68\% and 95\% CL.
In Ref.~\cite{Ellis:2018gqa}, a combined fit of EWPO, Higgs and diboson data was performed, marginalising over 20 operators in total, neglecting $\mathcal{O}(\Lambda^{-4})$ effects, at LO in the SMEFT.
We show in Fig.~\ref{fig:ellis-bounds} the marginalised 95\% CL bounds and central values obtained, marginalising over all operators.
A combined fit to EWPO, Higgs and diboson data was also performed in Ref.~\cite{Almeida:2018cld} to the 20 parameters of interest with 122 measurements. The results in the analysis are shown at both $\mathcal{O}(\Lambda^{-4})$ and $\mathcal{O}(\Lambda^{-2})$, in order to understand the impact of the higher order contributions. We show in Fig.~\ref{fig:eboli-bounds} the 95\% CL bounds obtained by marginalising over the operators.

\begin{figure}[t]
  \begin{center}
\includegraphics[scale=0.7]{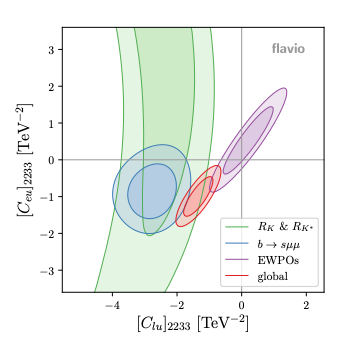}
    \caption{\small $1\sigma$ and $2\sigma$ likelihood contours for the $[C_{lu}]_{2233}$ and $[C_{eu}]_{2233}$ Wilson coefficients. The definitions of the operators given in Ref.~\cite{Aebischer:2018iyb}. Figure from Ref.~\cite{Aebischer:2018iyb}.
      \label{fig:smelli-bounds}
  }
  \end{center}
\end{figure}

We will finally turn to the flavour sector of the SMEFT. The \texttt{Python}  library smelli (SMEFT likelihood) ~\cite{Aebischer:2018iyb} provides the global likelihood for combined EWPO and flavour observables, with 265 measurements in total. 
As they combine sectors with typical scales above and below the electroweak scale, the RG running of the Wilson coefficients is taken into consideration.
We show in Fig.~\ref{fig:smelli-bounds} the $1\sigma$ and $2\sigma$ 2-dimensional likelihood contours for two Wilson coefficients considered in smelli which are of interest in top physics.

\paragraph{Summary}
We are in a position where fits to Wilson coefficients in the SMEFT are able to marginalise over many operators at a time. 
Furthermore, there has been a huge amount of progress in the SMEFT at NLO -- 
in addition to the top processes discussed above, there have also been many electroweak and Higgs processes computed through to NLO QCD in the SMEFT~\cite{Alioli:2018ljm,Artoisenet:2013puc,Maltoni:2013sma,Demartin:2014fia,Demartin:2015uha,Mimasu:2015nqa,Degrande:2016dqg}.
There has also been progress on fitting multiple sectors at once and with the full RGE at dimension-6 understood, in principle one should be able to combine measurements at different scales to constrain Wilson coefficients.
We are therefore now at a point where we can move towards truly global fits of the SMEFT.
%


\providecommand{\href}[2]{#2}\begingroup\raggedright\endgroup

\end{document}